\documentclass[aps,twocolumn,pra,superscriptaddress,amsmath,showpacs,tightenlines]{revtex4}
\usepackage{epsfig,graphicx,times}
\usepackage{amstext}
\usepackage{amsmath}            
\usepackage{amssymb}            
\usepackage{graphicx}           
\usepackage{latexsym}
\usepackage{bm}
\usepackage[colorlinks,citecolor=blue, linkcolor=blue,hyperindex,CJKbookmarks,dvipdfm]{hyperref}

\begin{document}

\title{Electromagnetially-induced-transparency-like ground-state cooling in a double-cavity optomechanical system}

\author{Yujie Guo}
\affiliation{Beijing Computational Science Research Center, Beijing 100084, China}
\affiliation{Department of Physics, Tsinghua University, Beijing 100084, China}
\author{Kai Li}
\affiliation{Beijing Computational Science Research Center, Beijing 100084, China}
\author{Wenjie Nie}
\affiliation{Beijing Computational Science Research Center, Beijing 100084, China}
\affiliation{Department of Physics, Tsinghua University, Beijing 100084, China}
\author{Yong Li}
\email{liyong@csrc.ac.cn}
\affiliation{Beijing Computational Science Research Center, Beijing 100084, China}
\affiliation{Synergetic Innovation Center of Quantum Information and Quantum Physics, University of Science and Technology of China, Hefei, Anhui 230026, China}

\date{\today}

\begin{abstract}
We propose to cool a mechanical resonator close to its ground state via an electromagnetically-induced-transparency- (EIT-) like cooling mechanism in a double-cavity optomechanical system, where an additional cavity couples to the original one in the standard optomechanical system. By choosing optimal parameters such that the cooling process of the mechanical resonator corresponds to the maximum value of the optical fluctuation spectrum and the heating process to the minimum one, the mechanical resonator can be cooled with the final mean phonon number less than that at the absence of the additional cavity. And we show the mechanical resonator may be cooled close to its ground state via such an EIT-like cooling mechanism even when the original resolved sideband condition is not fulfilled at the absence of the additional cavity.

\end{abstract}


\maketitle

\section{Introduction}

Cooling mechanical resonators (MRs) has become an important topic for various fields of physics~\cite{Review}.
It is a prerequisite to even get the ground-state cooling of MRs for their possible uses in quantum information processing~\cite{MR-Quantum}. Various experiments have demonstrated significant cooling of MRs in optomechanical systems~\cite{Cooling-Exp}. Recently, the ground-state cooling of MRs has already been achieved~\cite{Quantum-Drum,Teufel,Painter}.

So far, many theoretical cooling schemes~\cite{Electro-Mech-Cooling, Zhang2005PRL, Sideband-cooling, Sideband-cooling2, 3MC-Cooling, Xia2009PRL, Electro-Mech-EIT-Cooling, Other-OM-cooling,Genes, LiGaoxiang} have been proposed to achieve the ground-state cooling of MRs. Among them, the most studied and famous scheme is the (resolved) sideband cooling~\cite{Sideband-cooling} for a standard optomechanical system wherein the MR is coupled to the optical field via radiation pressure force. According to the quantum theory of sideband cooling of MRs~\cite{Sideband-cooling}, the desired fluctuation spectrum of the optical field that couples to the MR determines the transition rates of both cooling and heating processes of the MR, i.e. the fluctuation spectrum at the MR frequency $\omega_{m}$ causes the cooling transition, whereas the one at $-\omega_{m}$ causes the heating transition, corresponding to the anti-Stokes and the Stokes processes, respectively. In the resolved sideband case, the decay rate of the optical field (cavity field) is less than the frequency of the MR, that is, the (half-) width of the single Lorentzian peak of the optical fluctuation spectrum is less than the mechanical frequency, one may obtain the ground-state cooling of the MR by putting the cooling anti-Stokes process corresponding to the maximum value of the optical fluctuation spectrum and the heating Stokes process to a much smaller one.

However, except for some special optomechanical systems as in Ref.~\cite{Teufel,Painter}, the resolved sideband condition is hard to be fulfilled in many experimental optomechanical systems. Thus, other ground-state cooling schemes beyond the sideband cooling are required~\cite{Zhang2005PRL, Xia2009PRL, Electro-Mech-EIT-Cooling, Other-OM-cooling,Genes, LiGaoxiang}. Xia and Evers~\cite{Xia2009PRL} have applied the electromagnetically-induced-transparency- (EIT-) cooling scheme of the motion of trapped particle~\cite{EIT-cooling} to cool a MR when it couples to a three-level superconducting flux qubit. This EIT cooling works in the non-resolved sideband regime but suppresses the (carrier) heating processes by means of the EIT phenomenon~\cite{EIT} in three-level systems. The similar EIT-like cooling mechanism has been used to cool the MR when it couples to the single electronic spin qubit of nitrogen-vacancy impurity~\cite{Electro-Mech-EIT-Cooling}. Recently Genes \emph{et al.}~\cite{Genes} also proposed an EIT ground-state cooling scheme of MR via EIT in the three-level atomic medium in a hybrid optomechanical system.

Here, motivated by these works, we propose an EIT-like ground-state cooling scheme of MR in a double-cavity optomechanical system. In our model, the MR is coupled to the first one of the two coupled single-mode cavities (also called optical molecule~\cite{Jing2014}) via the radiation pressure force. The desired optical fluctuation spectrum to which the MR is subjected is determined by the two coupled cavities and splits from the single Lorentzian peak of the standard optomechanical case into two relatively narrower peaks with a dip emerging between them. When the decay rate of the second cavity is small enough (e.g., much smaller than that of the first one and the coupling strength between the cavities), the dip will be approximately close to zero and the corresponding spectrum will have the EIT-like form, similar to the EIT phenomenon in typical $\Lambda$-type three-level atomic systems~\cite{EIT}. By putting the cooling (anti-stokes) process of the MR corresponding to the maximum value of the optical fluctuation spectrum and the heating (Stokes) process to the minimum one, the mechanical resonator can be cooled better than that at the absence of the additional cavity, and even be cooled to the ground state.

Note that the EIT-like phenomenon in two coupled cavities has been achieved in experiments~\cite{ZhangJing}. The analog of the EIT-like phenomenon in coupled harmonic oscillators (e.g., bosonic cavity modes or mechanical resonators) with the EIT phenomenon of $\Lambda$-type three-level atomic systems has also been discussed~\cite{EIT-MR,Yusuf}. We also note that a ground-state cooling scheme was proposed in an optomechanical system involving two cavity modes and one MR~\cite{LiGaoxiang}. The difference between the work in Ref.~\cite{LiGaoxiang} and ours is the following: 1) In Ref.~\cite{LiGaoxiang}, two cavity modes are indirectly coupled to each other via the atomic medium in a mixed cavity system, here the two single-mode cavities are directly coupled in a photonic molecule system; 2) The MR couples to both the cavity modes in Ref.~\cite{LiGaoxiang}, while it couples to only one of the single-mode cavities in our model.


This paper is organized as follows. In Sec. II, we describe in detail our model  Hamiltonian and discuss the final mean phonon number analytically. In Sec. III, the detailed properties of the optical fluctuation spectrum via the EIT-like mechanism and the optimal cooling conditions are discussed. Finally, a brief conclusion is given in Sec. IV.

\section{Hamiltonian of the model and rate equations of the mechanical resonator}

\begin{figure}[ptbh]
\centering
\includegraphics[width=6.0cm]{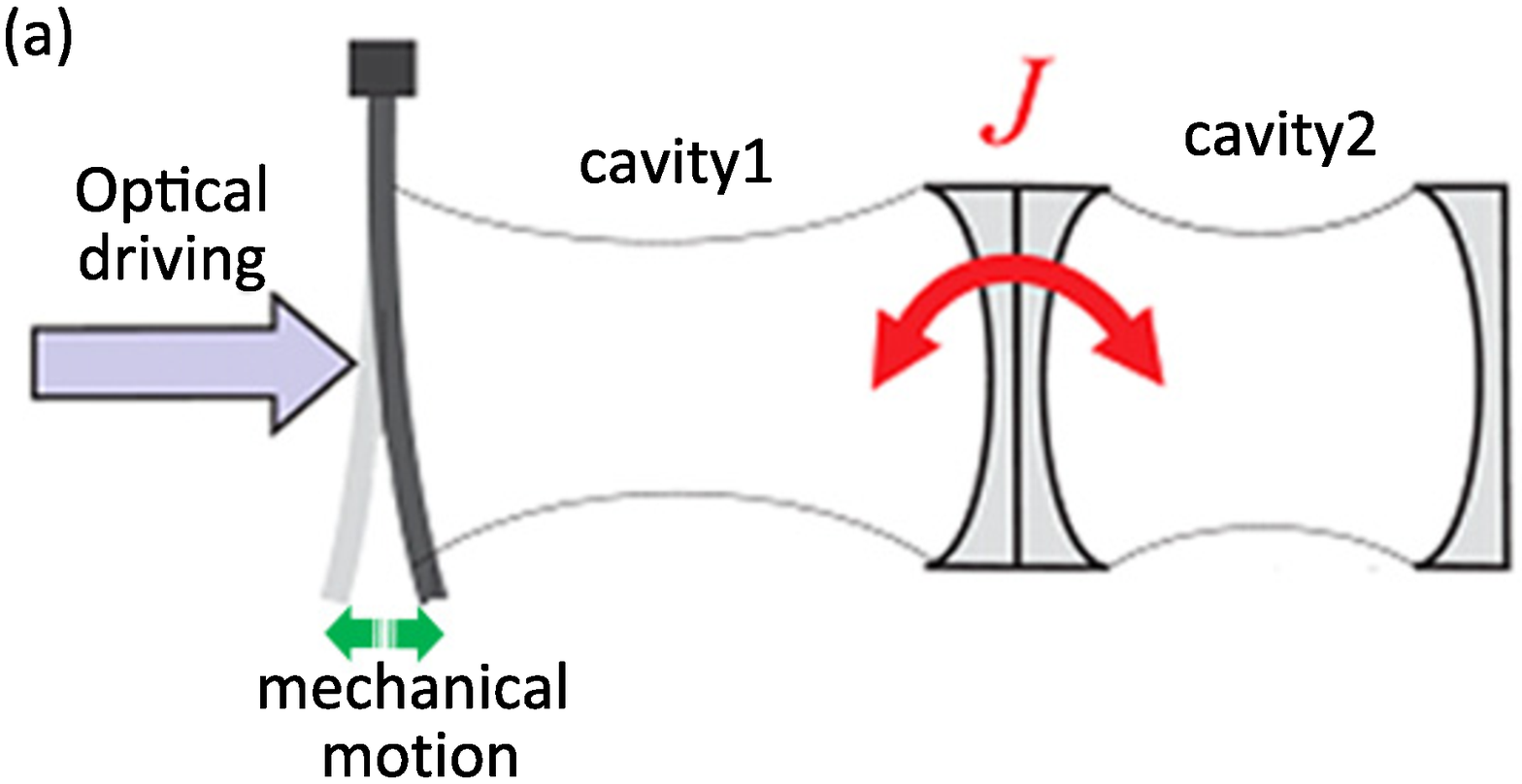}
\includegraphics[width=6.0cm]{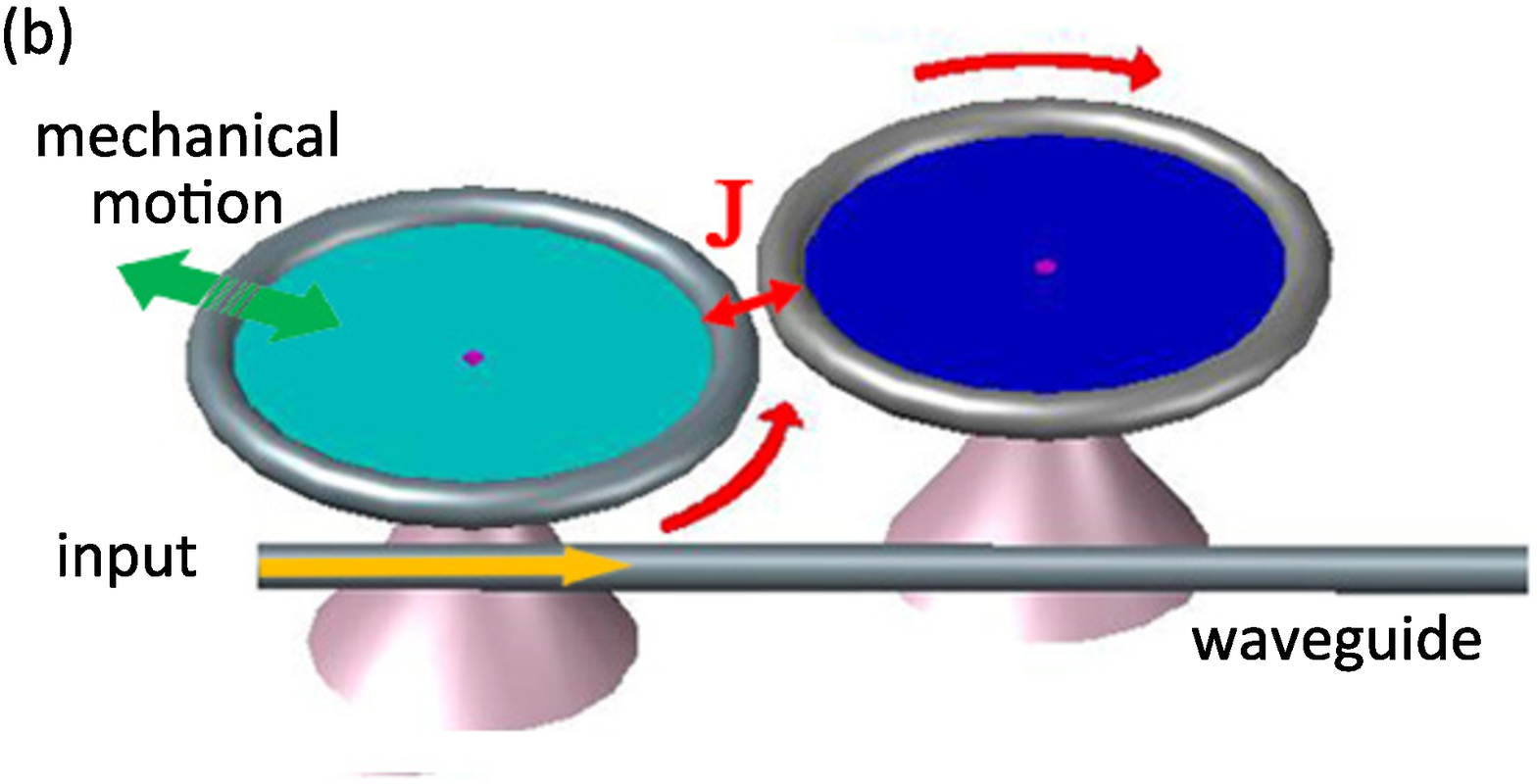}
\caption{(Color online) The schematic of the double-cavity optomechanical system with the possible realization in the system based on (a) Fabry-Perot cavities and (b) whispering gallery cavities.}%
\label{model}%
\end{figure}

The system we study here is composed of a MR and two coupled single-mode cavities. The MR couples to the first cavity which is driven by an external optical field, forming a standard optomechanical subsystem. The second cavity couples to the first one with the coupling strength $J$. In experiments, such a double-cavity optomechanical model can be achieved in the systems based on Fabry-Perot cavities or whispering gallery cavities~\cite{WGM} (see Fig.~\ref{model}.

The Hamiltonian of this system reads ($\hbar =1$)
\begin{align}
H  &  =\omega_{1}a_{1}^{\dag}a_{1}+\omega_{2}a_{2}^{\dag}a_{2}+J\left(  a_{1}^{\dag}a_{2}+a_{1}a_{2}^{\dag}\right) \nonumber\\
&  +\omega_{m}b^{\dag}b- g_{0}\left(  b^{\dag}+b\right)  a_{1}%
^{\dag}a_{1}\label{1}\\
&  +i \left(  \varepsilon a_{1}^{\dag}e^{-i\omega_{L}t}-\varepsilon^{\ast
}a_{1}e^{i\omega_{L}t}\right)  .\nonumber
\end{align}
Here $a_{1}$, $a_{2}$ and $b$ are the annihilation operators of the two cavity
modes and the MR, with $\omega_{1}$, $\omega_{2}$ and $\omega_{m}$ being their frequencies, respectively. $g_{0}$ is the single-photon optomechanical coupling coefficient. $\omega_{L}$ is the frequency of the driving field, and $\varepsilon$ is related to the power of the driving laser. Note that only the first cavity is driven by the external field and couples to the MR.

In a frame rotating at the driving frequency $\omega_{L}$, the Hamiltonian (\ref{1}) becomes
\begin{align}
H  &  =\Delta_{1}^{(0)}a_{1}^{\dag}a_{1}+\Delta_{2}a_{2}^{\dag}a_{2}+
J\left(  a_{1}^{\dag}a_{2}+a_{1}a_{2}^{\dag}\right) \nonumber\\
&  +\omega_{m}b^{\dag}b- g_{0}\left(  b^{\dag}+b\right)  a_{1}%
^{\dag}a_{1}\label{2}\\
&  +i \left(  \varepsilon a_{1}^{\dag}-\varepsilon^{\ast}a_{1}\right)
,\nonumber
\end{align}
where $\Delta_{1}^{(0)}=\omega_{1}-\omega_{L}$ and $\Delta_{2}=\omega_{2}-\omega_{L}$ are the detunings of the two cavity modes from the driving field, respectively.

By rewriting each operator as a sum of its steady state mean value and a zero-mean fluctuation like $a_{1}=\alpha_{1} +\delta a_{1}$,
$a_{2}=\alpha_{2} +\delta a_{2}$, $b= \beta  +\delta b$, and following the usual  linearization approach~\cite{Sideband-cooling,Sideband-cooling2} for the case of $\alpha_{1} \gg1$ in optomechanical systems, one can obtain the effective linearlized Hamiltonian of the fluctuation operators (hereafter we drop the notation ``$\delta$" for all the fluctuation operators for the sake of simplicity, like ``$\delta a_1 \rightarrow a_1$")
\begin{align}
H_{\rm{eff}}  &  ={\Delta}_{1} a_{1}^{\dag}a_{1}+\Delta_{2}%
a_{2}^{\dag}a_{2}+ J\left(  a_{1}^{\dag}a_{2}+a_{1}a_{2}^{\dag}\right)
\nonumber\\
&  +\omega_{m}b^{\dag}b- g\left(  b^{\dag}+b\right)  \left(
a_{1}^{\dag}+a_{1}\right)  , \label{3}%
\end{align}
where ${\Delta}_{1}=\Delta_{1}^{(0)}-g_{0} (\beta + \beta^{*})$ is the effective detuning of the first cavity mode, $g=g_{0}\alpha_{1} $ is the enhanced effective optomechanical coupling coefficient with the steady-state value
\begin{align}
\alpha_{1}&=\frac{\varepsilon}{\kappa_{1}+i{\Delta}_{1}
+\frac{J^{2}}{\kappa_{2}+i\Delta_{2}}},\nonumber\\
\alpha_{2} &=\frac{-iJ\alpha_{1}}{\kappa_{2}+i{\Delta}_{2}},\nonumber\\
\beta &=\frac{ig_{0}|\alpha_{1}|^{2}}{i\omega_{m}+\gamma_m}.
\end{align}
Without loss of generality, we have assumed the steady-state values $\alpha_{1}$ to be real. The last term in the second line of Eq.~(\ref{3}) describes the effective optomechanical coupling, where $a_{1}^{\dag}+a_{1}=:F$ represents the effective (dimensionless) optical force on the MR.

According to the effective Hamiltonian~(\ref{3}) and following the methods as given in Refs.~\cite{Sideband-cooling,Fermi}, one can write down the rate equations of the MR as
\begin{align}
\dot{P}_{n}  &  =\Gamma_{n\leftarrow n+1}P_{n+1}+\Gamma_{n\leftarrow
n-1}P_{n-1}\nonumber\\
&  -\Gamma_{n-1\leftarrow n}P_{n}-\Gamma_{n+1\leftarrow n}P_{n} \nonumber\\
&  +\gamma_{m}\left(  n_{m}+1\right)  \left(  n+1\right)  P_{n+1}+\gamma
_{m}n_{m}nP_{n-1}\nonumber\\
&  -\gamma_{m}\left(  n_{m}+1\right)  nP_{n}-\gamma_{m}n_{m}\left(
n+1\right)  P_{n} \label{4}
\end{align}
by eliminating the degrees of freedom of the optical field. Here, $P_{n}$ is the probability for the MR to be in the mechanical Fock state $\left\vert n\right\rangle $ with $n$ phonons. $\Gamma_{n^{\prime }\leftarrow n}$ represents the transition rate from the Fock state $\left\vert n\right\rangle $ to $\left\vert n^{\prime}\right\rangle $ induced by the effective optomechanical coupling, and by using Fermi's golden rule~\cite{Fermi}, one can obtain $\Gamma_{n\leftarrow n+1}=\left(  n+1\right) g^{2}S_{FF}\left(  \omega_{m}\right)  $ and $\Gamma_{n+1\leftarrow n}=\left(
n+1\right)  g^{2}S_{FF}\left(  -\omega_{m}\right)  $ with $S_{FF}\left(
\omega\right)  =\int dte^{i\omega t}\left\langle F\left(  t\right)  F\left(
0\right)  \right\rangle $ being the fluctuation spectrum of the optical force
$F=a_{1}^{\dag}+a_{1}$. The terms in the last two lines describe the
transition induced by the thermal bath, where $\gamma_{m}$ is the mechanical
damping rate, $n_{m}=\left(  e^{\hbar\omega_{m}/k_{B}T}-1\right)  ^{-1}$ is
the thermal phonon number with environment temperature $T$.

From the rate equations~(\ref{4}) one can solve the steady state final mean
phonon number of the mechanical resonator which reads
\begin{equation}
n_{f}=\frac{\gamma_{m}n_{m}+\gamma_{c}n_{c}}{\gamma_{m}+\gamma_{c}}, \label{5}%
\end{equation}
where
\begin{align}
\gamma_{c}  &  =g^{2}\left[  S_{FF}\left(  +\omega_{m}\right)  -S_{FF}\left(
-\omega_{m}\right)  \right]  ,\label{6}\\
n_{c}  &  =\frac{S_{FF}\left(  -\omega_{m}\right)  }{S_{FF}\left(  +\omega
_{m}\right)  -S_{FF}\left(  -\omega_{m}\right)  }. \label{cooling-limit}%
\end{align}
$n_{c}$ is the quantum limit of cooling, since $n_{f}\rightarrow n_{c}$ when
$\gamma_{m}\rightarrow0$. $\gamma_{c}$ is the so-called cooling rate. The
final mean phonon number $n_{f}$ is mainly determined by the positive and
negative frequency parts of the fluctuation spectrum, i.e. $S_{FF}\left(
\pm\omega_{m}\right)  $. Note that the positive frequency part $S_{FF}\left(
+\omega_{m}\right)  $ that relates to the transition rate $\Gamma
_{n\leftarrow n+1}$ determines the cooling process, whereas the negative
frequency part $S_{FF}\left(  -\omega_{m}\right)  $ that relates to
$\Gamma_{n+1\leftarrow n}$ determines the heating process. To cool the
mechanical resonator close to its ground state, we need to control the
fluctuation spectrum $S_{FF}\left(  \omega\right)  $ of the optical force,
i.e. strengthen the positive frequency part $S_{FF}\left(  +\omega_{m}\right)
$ and suppress the negative frequency part $S_{FF}\left(  -\omega_{m}\right)$. In other words, large cooling rate $\gamma_{c}$ and small cooling limit
$n_{c}$ are both required.

In the weak coupling regime, the reaction of the MR to
light can be neglected. So the fluctuation spectrum $S_{FF}\left(
\omega\right)  $ of the optical force $F=a_{1}^{\dag}+a_{1}$ is totally determined by the optical part in the effective Hamiltonian (\ref{3}):
\begin{align}
H_{\rm{op}}  = {\Delta}_{1}a_{1}^{\dag}a_{1}+ \Delta_{2}a_{2}^{\dag}%
a_{2}+ J\left(  a_{1}^{\dag}a_{2}+a_{1}a_{2}^{\dag}\right).  \label{Hop} \end{align}
Thus, $S_{FF}\left( \omega\right) $ can be easily obtained from the corresponding quantum Langevin equations
\begin{align}
\dot{a}_{1}  &  =-i{\Delta}_{1}a_{1}-iJa_{2}-\kappa_{1}a_{1}+\sqrt{2\kappa_{1}}a_{1,in},\nonumber\\
\dot{a}_{2}  &  =-i\Delta_{2}a_{2}-iJa_{1}-\kappa_{2}a_{2}+\sqrt{2\kappa_{2}}a_{2,in}, \label{7}%
\end{align}
where $\kappa_{1}$ and $\kappa_{2}$ are the cavity decay rates, $a_{1,in}$ and
$a_{2,in}$ are the noise operators with their non-zero correlation functions satisfying $\langle a_{j,in}\left(  t\right)  a_{j,in}^{\dag}\left(  t^{\prime
}\right)  \rangle =\delta\left(  t-t^{\prime}\right)  $
($j=1$, $2$). As a result, we obtain
\begin{equation}
S_{FF}\left(  \omega\right)  =\frac{1}{A\left(  \omega\right)  }+\frac
{1}{A^{\ast}\left(  \omega\right)  }, \label{8}%
\end{equation}
where $A\left(  \omega\right)  =\kappa_{1}-i\left(  \omega-{\Delta}%
_{1}\right)  +\frac{J^{2}}{\kappa_{2}-i\left(  \omega-\Delta_{2}\right)  }$.

\section{Optical fluctuation spectrum and EIT-like cooling}

As mentioned above, the cooling result of the MR is mainly determined by the positive and negative frequency parts of the optical fluctuation spectrum, i.e. $S_{FF}\left(  \pm\omega_{m}\right) $. In the following, we investigate the dependence of the fluctuation spectrum $S_{FF}\left(  \omega\right)  $ of the optical force on the parameters, e.g., the optical coupling coefficient between the two optical cavities, the (effective) optical detunings, in order to get the optimal cooling.

\begin{figure}[tbp]
\includegraphics[width=8.0cm]{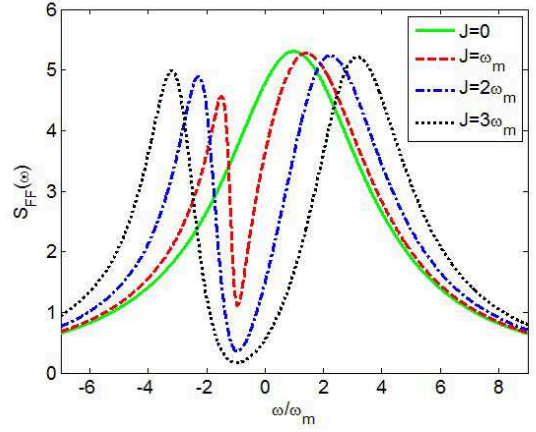}
\caption{(Color online) The optical fluctuation spectrum $S_{FF}\left(  \omega\right)  $ (in arbitrary units) as a function of the frequency $\omega $ with four different optical coupling coefficients $J$.
The effective detuning of the first cavity mode ${\Delta}_{1}=\omega_{m}$ and its decay rate is $\kappa_{1}=3\omega_{m}$. While the detuning of the
second cavity mode is $\Delta_{2}=-\omega_{m}$ and the corresponding decay rate is $\kappa_{2}=0.1\omega_{m}$.   }
\label{FIG2}
\end{figure}

At the absence of the second cavity ($J=0$), the profile of the optical fluctuation spectrum $S_{FF}\left(  \omega\right)$ has a Lorentzian shape with single peak located at $\omega={\Delta}_{1}$ and the (half-) width of the peak being $\kappa_{1}$. According to the sideband cooling mechanism~\cite{Sideband-cooling}, the necessary condition to get ground state cooling of the MR is that the sideband is resolved: $\kappa_{1}<\omega_{m}$. In the case of non-resolved sideband, the fluctuation spectrum values $S_{FF}\left(  +\omega_{m}\right)  $ and $S_{FF}\left(  -\omega_{m}\right)  $ determining respectively the cooling and heating processes are comparable and therefore the optimal cooling of the mechanical resonator is not achieved. Here, we focus on the case of non-resolved sideband for the first cavity $\left(  \kappa_{1}>\omega_{m}\right)$ in the double-cavity optomechanical system.

In Fig.~\ref{FIG2}, the optical fluctuation spectrum $S_{FF}\left(  \omega\right) $ versus the frequency $\omega $ is depicted with four different optical coupling coefficients. It is interesting that due to the existence of the coupling between two optical cavities ($J>0$), the single Lorentzian peak splits into two relatively narrower peaks and a dip emerges between them.
Physically the origin of the dip is similar to the two-photon resonance in the EIT phenomenon of three-level atomic system, as discussed in Ref.~\cite{Yusuf}. This means the tip of the spectrum $S_{FF}\left(  \omega\right) $ locates at $\omega = \Delta_{2}$).

Consequently, in order to suppress the heating process as much as possible, that is, to make the related fluctuation spectrum $S_{FF}(\omega= -\omega_{m})$ take the value of the tip, the corresponding optimal condition can be attained as
\begin{equation}
\Delta_{2}=-\omega_{m}.   \label{tip-position}
\end{equation}

In addition, it is found from Fig.~\ref{FIG2}, that the positions of the two peaks of fluctuation spectrum depend strongly on the optical coupling coefficient $J$. In order to maximize the transition rate of the cooling process, the fluctuation spectrum value $S_{FF}\left( \omega= +\omega_{m}\right)  $ determining the cooling process should be as large as possible. That is, we need to fix the centre of the right peak around $\omega = +\omega_{m}$.

In fact, in the double-cavity optomechanical system, these two new peaks
originate from the normal mode splitting, which can be seen by diagonalizing $H_{\rm{op}}$ in Eq.~(\ref{Hop}) [the optical parts of the effective Hamiltonian~(\ref{3})]
\begin{align}
H_{\rm{op}}= \Delta_{1}^{\prime}a_{1}^{\prime\dag}a_{1}^{\prime}+ \Delta
_{2}^{\prime}a_{2}^{\prime\dag}a_{2}^{\prime},
\end{align}
where
\begin{align}
\Delta_{1,2}^{\prime} &  =\frac{{\Delta}_{1}+\Delta_{2}}{2}\pm \sqrt
{J^{2}+\left(  \frac{{\Delta}_{1}-\Delta_{2}}{2}\right)  ^{2}}.
\end{align}
Here $a_{1}^{\prime}=a_{1}\cos\theta+a_{2}\sin\theta$ and $a_{2}^{\prime}=a_{1}\sin\theta-a_{2}\cos\theta$ are the annihilation operators for the diagonalized optical collective normal modes, where $\theta$ satisfies $\tan 2\theta ={2J}/({\Delta}_{1}-\Delta_{2})$. $\Delta_{1}^{\prime}$ ($\Delta_{2}^{\prime}$) is the eigen-frequency of diagonalized collective mode, corresponding to the location of the right (left) peak of the optical spectrum $S_{FF}(\omega)$. Thus, in order to maximize the transition rate of cooling process, the optimal cooling condition is that $S_{FF}(+\omega_{m})$ is just corresponding to the right peak, that is
\begin{equation}
\Delta_{1}^{\prime}=+\omega_{m}. \label{right-peak}
\end{equation}

Combining Eqs.~(\ref{tip-position}) and (\ref{right-peak}) one can obtain the optimal optical coupling coefficient as
\begin{equation}
J=\sqrt{2\omega_{m}\left(  \omega_{m}-{\Delta}_{1}\right)  }. \label{Optimal-J}
\end{equation}
When $\omega_{m} < {\Delta}_{1}$, the fact that the ``optimal" $J$ from Eq.~(\ref{Optimal-J}) means the right peak locates always at the right side of the point $\omega=+\omega_m$ and $S_{FF}(+\omega_{m}) < S_{FF}(\Delta_1^{\prime})$. Nevertheless, one can also get cooling of MR in this case. In this work, we focus on the case of $\omega_{m} > {\Delta}_{1}$.

\begin{figure}[tbp]
\includegraphics[width=8.0cm]{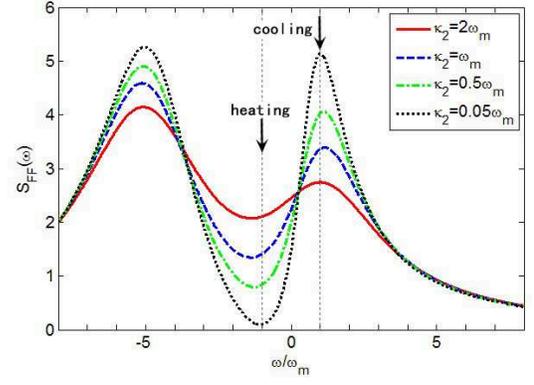}
\caption{(Color online) The optical fluctuation spectrum $S_{FF}\left(  \omega\right) $ (in arbitrary units) with four different decay rates $\kappa_{2}$ at the given optical effective detuning ${\Delta}_{1}=-3\omega_{m}$ and optimal detuning $\Delta_{2}=-\omega_{m}$ and corresponding optimal optical coupling coefficient $J=2\sqrt{2}\omega_{m}$. Here $\kappa_1=3\omega_m$.}
\label{FIG3}
\end{figure}

In Fig.~\ref{3}, the optical fluctuation spectrum $S_{FF}\left(  \omega\right) $ is depicted with four different decay rates $\kappa_{2}$ under the optimal conditions that $\Delta_{2}=-\omega_{m}$ and the value of optical coupling $J$ satisfies Eq.~(\ref{Optimal-J}). The effective detuning of the first cavity mode is selected as ${\Delta}_{1}=-3\omega_{m}$, which means and corresponding optimal optical coupling coefficient $J=2\sqrt{2}\omega_{m}$.

It is noted that even in the two optimal conditions with $\Delta_{2}=-\omega_{m}$ corresponding to minimal heating effect and $\Delta_{1}^{\prime }=+\omega_{m}$ corresponding to maximal cooling effect, we should also require the value of the tip of the related spectrum being close to zero in order to get the nice cooling, e.g., ground-state cooling. This can be obtained by selecting properly the decay rate of the second optical cavity $\kappa_{2}$ which determines the deep of the dip of the fluctuation spectrum. In fact, when $\kappa_2 $ is very small, e.g., $\kappa_2 \ll J $, the value of the tip is close to zero (as seen in Fig.~\ref{FIG2}). It is also clearly seen from Fig.~\ref{FIG3} that with decreasing the values of decay rates $\kappa_{2}$, the height of the peak increases while the dip gets close to zero gradually. This suggests that in the double-cavity optomechanical system a small decay rate $\kappa_{2}$ is preferable for the cooling of the mechanical resonator.

\begin{figure}[tbp]
\includegraphics[width=6.0cm]{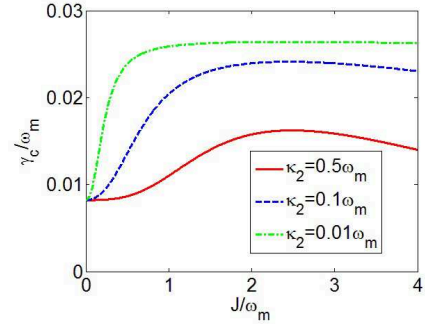}
\caption{(Color online) The cooling rate $\gamma_{c}$ as a function of the optical coupling coefficient $J$ with different decay rates $\kappa_{2}$. Here we fix $g=0.2 \omega_{m}$, and take $\Delta_{2}=-\omega_{m}$, $\kappa_1=3\omega_m$, and the optimal detuning ${\Delta}_{1}$ satisfying Eq.~(\ref{Optimal-J}).}
\label{FIG4}
\end{figure}
\begin{figure}[tbp]

\includegraphics[width=6.0cm]{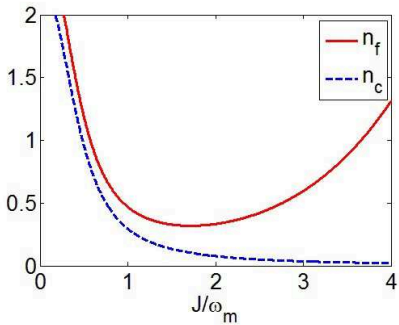}
\caption{(Color online) The mean phonon number $n_{f}$ as a function of the dimensionless optimal optical coupling coefficient $J$ with different decay rates $\kappa_{2}$. For the parameters, see the context.}
\label{FIG5}
\end{figure}

In Fig.~\ref{FIG4}, the cooling rate $\gamma_{c}$ is given as a function of the optical coupling coefficient $J$ with different decay rates $\kappa_{2}$ for the optimal conditions that $\Delta_{2}=-\omega_{m}$ and the detuning ${\Delta}_{1}$ satisfies Eq.~(\ref{Optimal-J}). We can know that the smaller $\kappa_2$ leads to better cooling rate. This agrees well with Eq.~(\ref{6}). The physical meaning is that the smaller $\kappa_2$ makes the dip of optical spectrum closer to zero, that is, better suppressing the heating process. The corresponding cooling limit in Eq.~\ref{cooling-limit}, $n_c$, for fixed $\kappa_{2}=0.1\omega_m$ is plotted in Fig.~\ref{FIG5} (the blue dash line). In principle, the cooling limit $n_c$ becomes more and more close to zero as increasing the coupling $J$. However, in a realistic system, the final mean phonon number, $n_f$, just takes the cooling limit $n_c$ when the MR thermal effect is much larger than the effect induced by the optical field, that is,
$\gamma_{m}n_{m} \ll \gamma_{c}n_{c}$ from Eq.~(\ref{5}). This will be not always valid, especially when $n_c\rightarrow 0$ for large $J$.

In order to consider the final cooling of the MR, we take a set of experimentally feasible parameters as follows~\cite{parameter}: $\omega_{m}=2\pi\times 20$ MHz, $Q_{m}=\omega_{m}/\gamma_{m}=8\times10^{4}$, $g_{0}=1.2\times10^{-4}\omega_{m}$, $|\varepsilon|=6000\omega_{m}$ (corresponding to the driving power $P\sim $ mW), and the initial phonon number $n_{m}=312$ (environment temperature $T=300$ mK). For the other parameters, we take the optimal optical detuning $\Delta_{2}=-\omega_{m}$, and the decay rates of the optical cavities $\kappa_{1}=3\omega_{m}$ and $\kappa_{2}=0.1\omega_{m}$. And note that here the effective detuning ${\Delta}_{1}$ always satisfies the optimal condition of Eq.~(\ref{Optimal-J}). With these parameters, the final mean phonon number is plotted in Fig.~\ref{FIG5} (the red solid line). One can see that the final mean phonon number $n_f$ can be less than $1$, e.g., $n_f \simeq 0.32 <1$ for $J=1.6 \omega_{m}$ where the corresponding ${\Delta}_{1}\simeq 0.12 \omega_{m}$ and $g \simeq 0.18 \omega_{m}$. That means even in the usual non-resloved sideband case (that is $\kappa_{1} > \omega_{m}$), the MR can be cooled close to its ground state due to the presence of the second cavity of good quality. The reason is that the interaction of the additional cavity with the first one makes the desired optical spectrum from the form of a Lorentian peak with the width larger than the MR's frequency (non-resolved sideband) to that with two peaks with the width of the right peak smaller than the MR's frequency. This means the effective resolved sideband condition is satisfied and thus the ground-state cooling of the MR can be achieved.

\section{Conclusion}

In summary, we have studied the cooling of a MR in a double-cavity optomechanical system. By applying the Fermi's golden rule approach to get the rate equation of the MR we obtain the analytic expression of the final mean phonon number of the MR in its steady state. Further, based on the EIT-like mechanism, we get the optimal cooling conditions by putting the cooling process of the MR corresponding to the right peak of the desired optical fluctuation spectrum and the heating process to the tip of the spectrum. We find the MR can be cooled close to its ground state via such an EIT-like cooling mechanism even when the original resolved sideband condition is not fulfilled without considering the additional cavity. Besides, the parameters we choose are experimentally feasible. This may benefit forward achieving quantum ground state of MRs in experiments and further possible applications involving quantum information processing based on MRs.

\begin{acknowledgments}
This work is supported by the NSFC (under Grant No. 11174027) and the National 973 program (under Grant No. 2012CB922104 and No. 2014CB921402).
\end{acknowledgments}

\end{document}